%% file: alpaka_p3ma_2017.tex
\documentclass{llncs}
\usepackage[utf8]{inputenc}
\usepackage[table]{xcolor}
\usepackage{tikz}
\usetikzlibrary{calc}
\usetikzlibrary{patterns}
\usepackage{pgfplots}
\usepackage{hhline}
\usepackage{numprint}
\npdecimalsign{.}
\nprounddigits{0}
\pgfplotsset{,colormap={mycolormap}{color(0cm)=(cyan); color(1cm)=(yellow); color(2cm)=(orange); color(3cm)=(red)}}
\usepackage{mathtools}
\usepackage{multirow}
\usepackage{xfrac}
\usepackage{pbox}
\usepackage{graphicx}
\usepackage{pifont}
\graphicspath{{./images/}}
\usepackage[group-separator={,}]{siunitx}

\definecolor{mygreen}{rgb}{0,0.6,0}
\definecolor{mygray}{rgb}{0.5,0.5,0.5}
\definecolor{mymauve}{rgb}{0.58,0,0.82}
\definecolor{myblue}{HTML}{E5F2FF}

\newcommand{\issupp}{\cellcolor{green!25}}
\newcommand{\nosupp}{\cellcolor{red!25}}
\newcommand{\lisupp}{\cellcolor{yellow!25}}

\usepackage{listings}
\definecolor{clr_haswell}{rgb}{0.1,0.1,0.7}
\definecolor{clr_haswell_double}{rgb}{0.1,0.5,0.5}
\definecolor{clr_k80}{rgb}{0.1,0.7,0.1}
\definecolor{clr_p100}{rgb}{0.1,0.5,0.5}
\definecolor{clr_p100_pcie}{rgb}{0.25,0.7,0.7}
\definecolor{clr_knl}{rgb}{0.7,0.1,0.7}
\definecolor{clr_power8}{rgb}{0.7,0.3,0.3}

\definecolor{clr_tbl_bg1}{rgb}{0.9,0.9,1.0}
\definecolor{clr_tbl_bg2}{rgb}{1.0,1.0,0.7}

\newcommand{\GFLOPSinner}{\sfrac{\text{GFLOPs}}{\text{s}}}
\newcommand{\GFLOPS}{$\GFLOPSinner{}$}
\newcommand{\TFLOPS}{$\sfrac{\text{TFLOPs}}{\text{s}}$}

\newcommand{\specialcell}[2][c]{%
  \begin{tabular}[#1]{@{}c@{}}#2\end{tabular}}
  
\begin{document}

\title{Tuning and optimization for a variety of many-core architectures without changing a single line of implementation code using the Alpaka library\thanks{
	This project has received funding from the European Unions Horizon 2020 research and innovation programme under grant agreement No 654220.
	This project received funding within the MEPHISTO project (BMBF-Förderkennzeichen 01IH16006C).
	Research leading to these results has in parts been carried out on the Human Brain Project PCP Pilot System JURON at the Juelich Supercomputing Centre, which received co-funding from the European Union (Grant Agreement no. 604102).
	We thank for the access to and support for the HPC cluster Taurus at the Centre for Information Services and High Performance Computing (ZIH), Technical University Dresden, as well as the cluster Hypnos at the Helmholtz-Zentrum Dresden -- Rossendorf.
}}

\author{ Alexander Matthes\inst{1,2}
    \and Ren\'e Widera\inst{1}
    \and Erik Zenker\inst{3}
    \and Benjamin Worpitz\inst{3}
    \and Axel Huebl\inst{1,2}
    \and Michael Bussmann\inst{1}
}
\institute{ Helmholtz-Zentrum Dresden -- Rossendorf
	\and Technische Universit\"at Dresden
	\and LogMeIn, Inc.
}
\maketitle
\begin{abstract}
We present an analysis on optimizing performance of a single C++11 source code using the Alpaka hardware abstraction library.
For this we use the general matrix multiplication (GEMM) algorithm in order to show that compilers can optimize Alpaka code effectively when tuning key parameters of the algorithm.
We do not intend to rival existing, highly optimized DGEMM versions, but merely choose this example to prove that Alpaka allows for platform-specific tuning with a single source code.
In addition we analyze the optimization potential available with vendor-specific compilers when confronted with the heavily templated abstractions of Alpaka.
We specifically test the code for bleeding edge architectures such as Nvidia‘s Tesla P100, Intel‘s Knights Landing (KNL) and Haswell architecture as well as IBM‘s Power8 system. On some of these we are able to reach almost 50\% of the peak floating point operation performance using the aforementioned means. When adding compiler-specific \lstinline|#pragma|s we are able to reach 5 \TFLOPS{} on a P100 and over 1 \TFLOPS{} on a KNL system.
\end{abstract}

\input{content/introduction.tex}
\input{content/design.tex}
\input{content/results.tex}
\input{content/analysis.tex}
\input{content/evaluation.tex}

\bibliography{citations}

\end{document}

%% file: content/introduction.tex
\section{Introduction}

\subsection{Motivation}

We have developed Alpaka~\cite{zenker2016alpaka} due to our own need in programming highly efficient algorithms for simulations~\cite{zenker2016performance} and data analysis on modern hardware in a portable manner.
The aim of our approach is to have a single C++ source code in which we can express all levels of parallelism available on modern compute hardware, using a parallel redundant hierarchy model similar to that found in CUDA or OpenCL.
Taking a look at the recent top ten high performance computing (HPC) systems~\cite{MSD+16}, it becomes clear that many-core architectures and heterogeneous systems are dominating the landscape and will continue to do so.

The main design goal of Alpaka is to describe all levels of parallelization available on modern heterogeneous hardware. It neither makes assumptions on the memory layout or access patterns, nor does it handle the underlaying resource and event management of the whole application, nor does it abstract the inter-node communication.

Our open-source projects PIConGPU~\cite{PIConGPU2013,burau2010picongpu} and HaseOnGPU~\cite{eckert2016haseongpu} both use Alpaka for the kernel abstraction for various many-core hardware~\cite{zenker2016performance,zenker2016alpaka}, but different libraries for the mentioned topics not handled by Alpaka, like
Graybat~\cite{Zen16} for the network communication, mallocMC for the memory management or \mbox{libPMacc} for containers and asynchronous event handling.
Alpaka is not meant as full grown solution for developing or porting whole HPC applications, but as a single-purpose library that can easily be included into the individual software of an exiting HPC project.
We have chosen to provide a lightweight, yet powerful C++ meta programming library for coherently expressing parallelism for a large variety of many-core platforms.

Modern C++11 template programming enables us to implement an abstraction layer between the application and the various, often vendor-specific programming models available for programming many-core hardware. With modern compilers the abstraction layer is completely resolved during compilation, leaving only efficient code in the binary.

While performance portability and close-to-zero overhead of Alpaka code could be shown in previous work \cite{zenker2016alpaka} we will here concentrate on a subject important for high performance computing, namely optimization of code for various hardware platforms by means of tuning and vendor-specific compiler optimizations while maintaining a single-source, portable code. The presence of architecture independent parameters outside the algorithm implementation itself may also enable auto-tuning in a later step.

We will show that indeed parameter tuning and compiler optimization generate highly efficient code on various platforms. However, we will discuss some pitfalls of this approach that arise due to limiting tuning parameters to a small number and due to the lack of full support for C++11 in some vendor compilers.

\subsection{Alpaka}

\begin{figure}[tb]
\resizebox{1\textwidth}{!}{\includegraphics{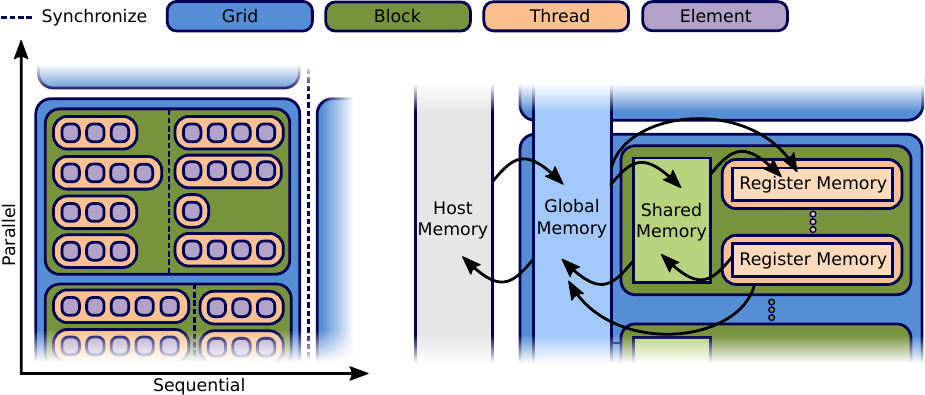}}
\caption{Systematic view of the abstract redundant parallel hierarchy model of Alpaka taken from~\cite{zenker2016alpaka}. A compute device works on a grid, inside this grid are blocks. Every block has the same amount of threads, which are defined by the user as kernels. Explicit looping over elements inside the kernel enables autovectorization but also gives a better utilization for simple kernels. For each computation layer Alpaka introduces an appropriate memory layer. The copies between those are explicit (depicted as arrows).}
\label {fig:alpaka-model}
\end{figure}

Alpaka allows for a multidimensional, hierarchical abstraction of computation hardware as seen in Fig. \ref{fig:alpaka-model}. Kernels are written and run as threads executed in a task parallel manner. Threads are organized in \emph{blocks}, which themselves are organized in \emph{grids}. Every thread inside a block is assumed to run in parallel to the other threads in the same block, enabling intra-block synchronization. Blocks on the other hand may run concurrently or sequentially inside a grid. Every execution layer has a corresponding memory hierarchy level.
In addition to task-parallel execution Alpaka introduces an element layer inside the thread level for data-parallel execution, where the same instruction or program is executed for multiple data. This latter level is usually used for expressing vectorization.

For any given hardware, these layers are mapped onto the hardware using a suitable back end. As such, Alpaka does not implement any functionality beyond this mapping and the underlying optimizations come form the back end and mapping chosen for a specific hardware.

Alpaka currently supports Nvidia's CUDA~\cite{CudaPG}, OpenMP~\cite{dagum1998openmp} 2 and 4, Boost Fibers and C++Threads as back ends. Furthermore, we have started to add Open\-ACC~\cite{Ope15} and Thread Building Blocks~\cite{Int17} (TBB) support, while support for AMD HIP~\cite{AMD15} is foreseen for the near future. Alpaka has two accelerators using OpenMP 2: One is running blocks in a grid concurrently, the other one threads inside a block. For the first one only one thread per block is allowed. With the same constraint it is possible to run the code sequentially with Alpaka.

In the scope of this paper we will restrict ourselves to the OpenMP 2 Blocks and Nvidia CUDA back ends so that we are able to compare our new results to our previous work.
Although OpenCL~\cite{OpenCL15} is widely supported, it is not suitable as Alpaka back end, as it is not single source C++. SYCL~\cite{KOWG15,WMA+16} has the goal to close this gap and will probably be considered in the future. C++ AMP~\cite{MC13} looks similarly promising, but fails in support of current HPC architectures.

Alpaka leaves performance enhancements due to data layout to the user or another, independent library. Memory in Alpaka thus is always represented by a plain pointer. This strategy leaves room for optimization, but currently requires more development effort by the user.

Optimized memory access patterns are as important for achieving performance as expression of algorithmic parallelism and we have carefully chosen the example GEMM algorithm as it is simple enough to go without memory abstraction. However, optimizing memory access and memory copies is outside the scope of Alpaka, which distinguishes our approach from the design goals of libraries such as Kokkos~\cite{edwards2013kokkos} or RAJA~\cite{hornung2014raja} that aim for providing a full software 
environment for portable programming of many-core architectures. A separate memory access abstraction library is planned, but will be an independent, orthogonal part of the already mentioned software stack.

%% file: content/design.tex
\section{The Alpaka general matrix multiplication implementation}

Similar to~\cite{zenker2016alpaka} we use a general matrix multiplication (GEMM) example
\begin{equation}
C = \alpha \cdot A \cdot B + \beta \cdot C
\end{equation}
for performance tuning, as it allows for tiling without the need for changing the memory representation of the matrices.

For the sake of simplicity we choose $A$, $B$ and $C$ to be quadratic matrices with $N$ rows and columns each. The total number of floating point operations then follows as
\begin{equation}
\text{O}(N) = 3 N^2 + 2 N^3 \approx 2 N^3 \enspace . \label{eq:O_of_N}
\end{equation}

The number of elements per thread $e$ and threads per block $t$ result in the number of blocks in the grid
\begin{equation}
\text{B}(e,t) = {N \over {t \cdot e}} \enspace ,
\end{equation}
whereby $t=1$ for the OpenMP 2 Blocks and the sequential accelerator.

We measure the time $t$ in seconds for the run of the algorithm without copy operations to device memory, keeping the maximum over ten runs. With this we calculate the performance $P$ in \GFLOPS{} as

\begin{equation}
\text{P}(N,t) = \frac{\text{O}(N)}{t} \cdot 10^{-9} = \frac{2 N^3}{t} \cdot 10^{-9} \enspace . \label{eq:P_of_N}
\end{equation}

\subsection{Tiled GEMM algorithm}


\begin{figure}[tb]
\def\svgwidth{\textwidth}
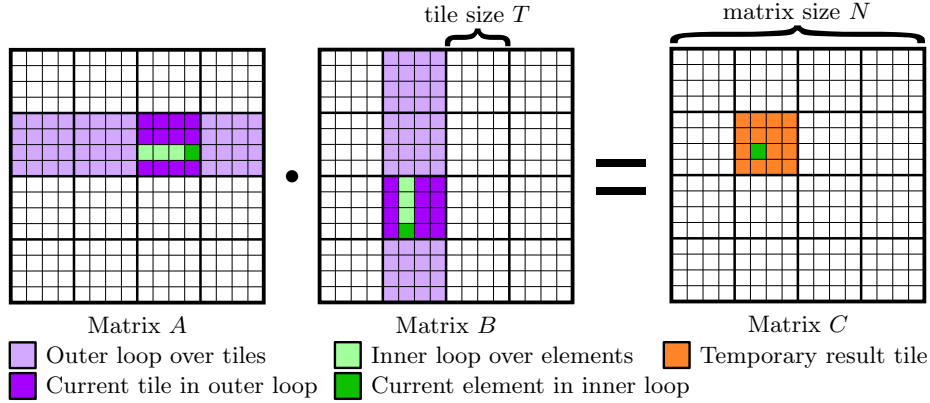
\caption{Performance critical $A \cdot B$ part of the GEMM using a tiling strategy. A thread iterates over smaller sub matrices (tiles) in $A$ and $B$ (purple), performs the matrix multiplication per tile using the element layer (green) for vectorization, and adds it to a temporary thread local $C$ tile (orange). The remaining part of the GEMM algorithm using the temporary $C$ tile needs to load and write the $C$ matrix only once (streaming), thus it doesn't need to be cached.}
\label {fig:gemm-alg}
\end{figure}

There exist many highly optimized GEMM implementations, reaching up to 90~\%~\cite{li2012optimized} of the theoretical peak performance.
The solution depicted here is not intended to compete with these algorithms.
Instead, it serves as a code example reasonably designed to exploit parallelism on many-core hardware.
As such, it already achieves 20 \% of the peak performance without tuning, which is a value regularly found in applications.
In the following, we aim to show that Alpaka allows for platform specific tuning by parameter tuning without specializing the implementation.
As long as the tiles of the two matrices $A$, $B$ fit completely in the cache memory, increasing the tile size will usually result in better performance. Based on the size $S$ in bytes of the data type used (single or double precision) the required cache size $K$ is
\begin{equation}
\text{K}(S,T) = 2 T^2 S \enspace . \label{eq:tilesize}
\end{equation}

The tiling matrix multiplication has $N_\text{blocks} = \sfrac{N}{T}$ tiles in each matrix dimension. For every tile of $C$ $N_\text{blocks}$ tiles of $A$ and $B$ need to be loaded (see Fig. \ref{fig:gemm-alg}). Furthermore the $C$ tile itself needs to be loaded, leading a total number of

\begin{equation}
\text{M}(N,T) = N_\text{blocks}^2 (2 T^2 N_\text{blocks} + T^2) = 2 \frac{N^3}{T} + N^2 = N^2 \bigg( 2\frac{N}{T} + 1 \bigg) \label{M_of_N_T}
\end{equation}
memory operations, which gives us the ratio of compute to memory operations as

\begin{equation}
\text{R}(N,T) = \frac{\text{O}(N)}{\text{M}(N,T)} = \frac{2 N^3}{N^2 (2\frac{N}{T} + 1)} = \frac{2 N}{(2 \frac{N}{T} + 1)} = \frac{2 N}{\frac{2 N+T}{T}} = \frac{2NT}{2N+T} \label{eq:R_N_T}
\end{equation}
with $\lim\limits_{N \to \infty}\text{R}(N,T) = T$, showing again that larger tile sizes are preferable

With cache hierarchies present in most modern architectures, it is not trivially predictable for which cache $T$ should be optimized.
We thus chose to calculate one tile of the matrix $C$ per Alpaka block. Every element in the block calculates one entry in the $C$ tile. We use a two dimensional indexing for the parallelization levels. Every element stores the partial result of $\alpha \cdot A \cdot B$ in element local memory.
Depending on the architecture, we can increase the number of elements per block by increasing the number of threads per block, which makes sense for GPUs, or the number of elements per thread, which should enable autovectorization for CPUs.

\begin{figure}[tb]
\begin{lstlisting}[basicstyle=\scriptsize\ttfamily,caption={Settings for the tiled matrix multiplication. \texttt{OptimalVectorSize::}\allowbreak{}\texttt{type::}\allowbreak{}\texttt{value} represents the tile size $T$. The parameters and the loop \textcolor{green!40!black}{\texttt{\#pragma}}s can directly be used inside the kernel.}, label={lst:parameters}, escapeinside={|}{|}]
// Class for optimal tile size depending on the Accelerator type
template< typename T_Acc >
struct OptimalVectorSize {
    using type = alpaka::dim::DimInt<1u>;
};
// Number of elements per tiles predefined, but changeable as compiler option
#ifndef GPU_ELEM_NUM
    #define GPU_ELEM_NUM 2u
#endif
#ifndef OMP_ELEM_NUM
    #define OMP_ELEM_NUM 256u
#endif
// Specialization of the tile size type for CUDA, steered by GPU_ELEM_NUM
#ifdef ALPAKA_ACC_GPU_CUDA_ENABLED
    template< typename... T_Args >
    struct OptimalVectorSize< alpaka::acc::AccGpuCudaRt< T_Args... > > {
        using type = alpaka::dim::DimInt<GPU_ELEM_NUM>;
    };
#endif
// Specialization for OpenMP Blocks, steered by OMP_ELEM_NUM
#ifdef ALPAKA_ACC_CPU_B_OMP2_T_SEQ_ENABLED
    template< typename... T_Args >
    struct OptimalVectorSize< alpaka::acc::AccCpuOmp2Blocks< T_Args... > > {
        using type = alpaka::dim::DimInt<OMP_ELEM_NUM>;
    };
#endif
// Easily extensible macro for every independent loop
#define VECTOR_PRAGMA \
    _Pragma ("ivdep") \
    _Pragma ("GCC ivdep")
\end{lstlisting}
\end{figure}

We implement the tile size $T$ as an accelerator dependent class as seen in Listing \ref{lst:parameters}, thus avoiding mixing tuning and kernel parameters.
It is set via \lstinline|#define|, thus making tuning easier.
The matrix sizes $N$ are passed as kernel parameters (not shown).

\subsection{Architectures}

\begin{table}[tb]
\centering
\bgroup
\def\arraystretch{1.15}%
\setlength\tabcolsep{2pt}
\scriptsize
\begin{tabular}{ccccc}
\hline 
\multicolumn{2}{c}{Vendor} & \multicolumn{3}{c}{ Nvidia } \\ 
\hline 
\multicolumn{2}{c}{Architecture} & K80 & \multicolumn{2}{c}{ P100 } \\ 
\hline
\multicolumn{2}{c}{Interconnect to host} & PCIe & nvlink & PCIe \\ 
\hline 
\multicolumn{2}{c}{Number of SMs} & 13 \cite{NC14} & \multicolumn{2}{c}{ 56 \cite{NC16} } \\
\hline 
\multirow{2}{*}{Cores per SM}  & SP & 192 \cite{NC14} & \multicolumn{2}{c}{ 64 \cite{NC16} } \\ \cline{2-5}
 & DP & 64 \cite{NC14} & \multicolumn{2}{c}{ 32 \cite{NC16} } \\ 
\hline 
\multicolumn{2}{c}{Shared memory per SM} & 112 KB \cite{NC14} & \multicolumn{2}{c}{ 48 KB  \cite{NC16} } \\
\hline 
\multicolumn{2}{c}{Registers per SM (32 Bit)} & \multicolumn{3}{c}{ \num{131072} \cite{NC14} \cite{NC16} } \\
\hline 
\multicolumn{2}{c}{\multirow{2}{*}{Clock frequency}} & $0.88$ Ghz & \multirow{2}{*}{$1.48$ Ghz} & \multirow{2}{*}{$1.39$ Ghz} \\ 
\multicolumn{2}{c}{} & (Boost clock) & & \\ 
\hline 
Theoretical peak & SP & $4.37$ \TFLOPS{} \cite{Nvi14} & $10.6$ \TFLOPS{} \cite{Nvi16} & $9.3$ \TFLOPS{} \cite{Nvi16} \\ \cline{2-5}
performance & DP & $1.46$ \TFLOPS{} \cite{Nvi14} & $5.3$ \TFLOPS{} \cite{Nvi16} & $4.7$ \TFLOPS{} \cite{Nvi16} \\
\hline 
\multicolumn{2}{c}{Release date} & Q4/2014 & \multicolumn{2}{c}{Q4/2016} \\ 
\hline 
\end{tabular}
\egroup
\vspace{2pt}
\caption{Single (SP) and double (DP) precision peak performances and other characteristic variables of GPUs considered in this paper. Notice that the P100 connected via nvlink has a higher frequency and thus a higher theoretical peak performance. The K80 has two GPU chips on one board of which we use only one. The cores of GPUs are grouped in Streaming Multiprocessors (SMs) similar to CPU sockets.}
\label{tab:gpus}
\end{table}

We test Nvidia K80 and P100 GPUs. The K80 and the PCIe version of the P100 are hosted in the cluster Hypnos at the Helmholtz-Zentrum Dresden -- Rossendorf whereas an nvlink using version of the P100 is part of the OpenPower pilot system JURON at the Jülich Supercomputing Center. All GPU architectures considered in this paper are listed in Tab.~\ref{tab:gpus}.

As modern GPUs can directly access host CPU memory, we test both manual offloading and Nvidia unified memory. For the first case we do not measure the time for explicit memory transfer between CPU and GPU. Be aware that memory handling is not part of Alpaka and native vendor code is used when necessary. We thus focus on measuring algorithmic performance while disregarding analysis of e.g. efficient latency hiding when offloading code to an accelerator.

\begin{table}[tb]
\centering
\bgroup
\def\arraystretch{1.15}%
\setlength\tabcolsep{2pt}
\scriptsize
\begin{tabular}{ccccc}
\hline 
\multicolumn{2}{c}{\multirow{2}{*}{Vendor and architecture}} & Intel Xeon & Intel Xeon Phi™ & IBM \\ 
\multicolumn{2}{c}{} & E5-2680 v3 (Haswell)& CPU 7210 (KNL) & Power8 \\ 
\hline 
\multicolumn{2}{c}{Used sockets} & 2 & 1 & 2 \\ 
\hline 
\multicolumn{2}{c}{Total number of cores $n$} & 24 & 64 & 20 \\ 
\hline 
\multicolumn{2}{c}{Hardware threads per core} & 1 & 4 & 8 \\ 
\hline 
\multicolumn{2}{c}{\multirow{2}{*}{Clock frequency $f$}} & $2.1$ Ghz (AVX & \multirow{2}{*}{$1.3$ Ghz} & \multirow{2}{*}{$4.02$ Ghz} \\ 
\multicolumn{2}{c}{} & base frequency \cite{New14}) & & \\ 
\hline 
FLOP per cycle & SP & 64 $_{(2 \cdot \text{AVX}, \text{FMA})}$ & 128 $_{(2 \cdot \text{AVX-512}, \text{FMA})}$ & 16 \cite{Her16} \\ \cline{2-5}
and core $o$ & DP & 32 $_{(2 \cdot \text{AVX}, \text{FMA})}$ & 64 $_{(2 \cdot \text{AVX-512}, \text{FMA})}$ & 8 \cite{Her16} \\
\hline 
Theoretical peak & SP & $1.61$ \TFLOPS{} & $5.33$ \TFLOPS{} & $1.29$ \TFLOPS{} \\ \cline{2-5}
performance (\ref{eq:peak_performance}) & DP & $0.81$ \TFLOPS{} & $2.66$ \TFLOPS{} & $0.64$ \TFLOPS{} \\
\hline 
Cache sizes & L1 & \multicolumn{3}{c}{$64$ KB (core)} \\ \cline{2-5}
reducing the & L2 & $256$ KB (core)& $1$ MB (2 cores) & $512$ KB (core) \\ \cline{2-5}
memory latency & L3 & $30$ MB (socket) & -- & $80$ MB (socket) \\
\hline
\multicolumn{2}{c}{Release date} & Q3/2014 & Q2/2016 & Q2/2014 \\ 
\hline 
\end{tabular}
\egroup
\vspace{2pt}
\caption{Single (SP) and double (DP) precision theoretical peak performances (see Eq. \ref{eq:peak_performance}) and other characteristic variables of CPUs considered in this paper. Performance gains come mostly from vector operations and fused multiply adds, especially for Intel CPUs, and higher clock frequencies when running on Power8.}
\label{tab:cpus}
\end{table}

Intel Xeon E5-2680 v3 (Haswell) and Xeon Phi Knights Landing (KNL) architectures are hosted on the HPC cluster Taurus located at Technical University Dresden whereas the Power8 processor is also part of the HPC pilot system JURON. The CPU architectures considered in this paper are listed in Tab.\ref{tab:cpus}.

Clock frequency $f$, FLOP per cycle and core $o$, and number of cores $n$ give the theoretical peak performance
\begin{equation}
\text{P}(f,o,n) = f \cdot o \cdot n \enspace . \label{eq:peak_performance}
\end{equation}

The Haswell CPU does not have hyperthreading activated and has two AVX units per core, which allows for instruction level parallelism and thus up to 64 single precision floating point operations (FLOPs) per cycle and clock. For measurements we use 2 sockets resulting in a total amount of 24 cores.
The KNL architecture allows for up to 128 single precision floating point operations per cycle and core. With hyperthreading activated this architecture can be used similar to a multi-core CPU with 256 independent threads.
The IBM Power8 processor has a uniquely high CPU frequency of 4 Ghz, but the lowest peak performance of all tested systems. However, with 8 hardware threads per core, 160 independent tasks can be executed without a context switch, allowing for high levels of parallelism.

\begin{table}[tb]
\centering
\bgroup
\def\arraystretch{1.15}%
\setlength\tabcolsep{2pt}
\scriptsize
\begin{tabular}{ccccc}
\hline 
 & Intel Compiler & CUDA & XL Compiler & GNU Compiler \\
\hline
Haswell & \issupp\specialcell{\lstinline|-Ofast -xHost| \\ (Version: 17.0.0)} & \nosupp-- & \nosupp-- & \issupp\specialcell{\lstinline|-Ofast -mtune=native| \\ \lstinline|-march=native| \\ (Version: 6.2)} \\
\hline
KNL & \issupp\specialcell{\lstinline|-Ofast -xHost| \\ (Version: 17.0.0)} & \nosupp-- & \nosupp-- & \issupp\specialcell{\lstinline|-Ofast| \\ \lstinline|-mtune=native| \\ \lstinline|-march=native| \\ (Version: 6.2)} \\
\hline
Tesla P100 & \nosupp-- & \issupp\specialcell{\lstinline|--use_fast_math| \\ (Version: 8.0.44)} & \nosupp-- & \lisupp\specialcell{\lstinline|-mtune=native -march=native| \\ (Version: 5.3, only host)} \\
\hline
Tesla K80 & \nosupp-- & \issupp\specialcell{\lstinline|--use_fast_math| \\ (Version: 8.0.44)} & \nosupp-- & \lisupp\specialcell{\lstinline|-mtune=native -march=native| \\ (Version: 5.3, only host)} \\
\hline
Power8 & \nosupp-- & \nosupp-- & \issupp\specialcell{\lstinline|-O5| \\ (Version: 14.01) \\ \textbf{(Only for C!)}} & \issupp\specialcell{\lstinline|-Ofast -mtune=native| \\ \lstinline|-mcpu=native -mveclibabi=mass| \\ (Version: 6.3)} \\
\hline

\hline 
\end{tabular}
\egroup
\vspace{5pt}
\caption{Compilers, compiler options, and compiler versions considered for every architecture in this paper. Every binary is compiled on the same system it is run on later, allowing for architecture- and system-aware compiler optimization.}
\label{tab:compilers}
\end{table}

We test different compilers for most architectures, see Tab.~\ref{tab:compilers}. The GNU compiler is used as a reference available for all architectures and for GPUs to compile the steering host code.
\subsection{Single source code file vs. optimization}

As pointer alignment and dependencies cannot be known at compile time, autovectorization needs some hints from developer side.
As pointed out, applications or additional libraries can provide additional information on data types that fosters autovectorization when using Alpaka.
We thus are forced to add compiler dependent \lstinline|#pragma|s, namely \lstinline|#pragma ivdep| and \lstinline|#pragma GCC ivdep| for the Intel and GNU C++ compilers, respectively, in order to declare pointers inside loops as independent and executable in parallel.
Furthermore, all memory is aligned to a multiplier of 64 with \lstinline|__declspec|\allowbreak{}\lstinline|(align(64))| (Intel) and \lstinline|__attribute__ ((aligned (64)))| (GNU compiler), which makes it faster to load whole chunks of memory to vector registers on some architectures. As one cannot pass this information via function parameters, we also explicitly tell the compilers about this in the most time critical loop over the $A$ and $B$ tiles with \lstinline|__assume_aligned| (Intel) and \lstinline|__builtin_assume_aligned| (GNU).

\subsubsection*{XL C++ work around}

Alpaka is a very demanding C++ code and most compilers fully support C++11, with the exception of the IBM XL compiler. For this reason we move the most performance critical code, the matrix multiplication of  tiles in $A$ and $B$, to an extra C file for every XL test and compile all C code with the XL compiler, while the C++ code including all Alpaka abstractions is compiled with the GNU C++ compiler. This means that we are not testing XL's OpenMP implementation. With full C++11 support by the IBM compiler we expect similar to better performance than we see with this approach.
This workaround currently breaks our single source goal and prevents code optimizations like code inlining, but still helps to improve performance compared to using just the GNU compiler.

\subsubsection{KNL specific parameter settings}
The Intel KNL is programmable similarly to a CPU, but like an offloading acceleration device it brings its own dedicated memory called MCDRAM. Compared to the global RAM the latency is almost the same, but the bandwidth around five times higher with over $450 \  \sfrac{\text{GB}}{\text{s}}$ (\cite{JRS16},~p.~20).
The Intel KNL supports three modes of accessing the MCDRAM: As a cache for RAM, directly accessed (flat memory) or a hybrid mode, where a part is used as cache and another part as flat memory. The first two modes are compared in performance, as they form opposite cases.
The Intel KNL can furthermore be operated in different cluster modes, which may improve the cache latency. In this paper we restrict ourselves to using quadrant mode only.

\subsubsection{Multidimensional parameter tuning}

We choose $T$ and the number of hardware threads as tuning parameters before running scaling tests for different matrix sizes $N$. Tuning is performed for a fixed $N=10240$ as a good compromise between runtime and problem size and further for an arbitrary $N=7168$, thus avoiding effects only occurring at some certain combinations of parameters.
After finding optimal parameter sets scaling tests with matrix sizes from $N=1024$ up to $N=20480$ with an increment of $\Delta N=1024$ are performed.
We repeat every measurement first 5 than 10 times, which in all cases yield the same maximum result.
This shows that any effects visible are not due to statistics, and we thus refrain from averaging over more measurements.

%% file: images/tiling-alg.pdf_tex
\begingroup%
  \makeatletter%
  \providecommand\color[2][]{%
    \errmessage{(Inkscape) Color is used for the text in Inkscape, but the package 'color.sty' is not loaded}%
    \renewcommand\color[2][]{}%
  }%
  \providecommand\transparent[1]{%
    \errmessage{(Inkscape) Transparency is used (non-zero) for the text in Inkscape, but the package 'transparent.sty' is not loaded}%
    \renewcommand\transparent[1]{}%
  }%
  \providecommand\rotatebox[2]{#2}%
  \ifx\svgwidth\undefined%
    \setlength{\unitlength}{544.09110963bp}%
    \ifx\svgscale\undefined%
      \relax%
    \else%
      \setlength{\unitlength}{\unitlength * \real{\svgscale}}%
    \fi%
  \else%
    \setlength{\unitlength}{\svgwidth}%
  \fi%
  \global\let\svgwidth\undefined%
  \global\let\svgscale\undefined%
  \makeatother%
  \begin{picture}(1,0.43282423)%
    \put(0,0){\includegraphics[width=\unitlength,page=1]{tiling-alg.pdf}}%
    \put(0.14032038,0.07264772){\color[rgb]{0,0,0}\makebox(0,0)[lb]{\smash{\makebox[0pt]{Matrix $A$}}}}%
    \put(0,0){\includegraphics[width=\unitlength,page=2]{tiling-alg.pdf}}%
    \put(0.47687229,0.07286027){\color[rgb]{0,0,0}\makebox(0,0)[lb]{\smash{\makebox[0pt]{Matrix $B$}}}}%
    \put(0,0){\includegraphics[width=\unitlength,page=3]{tiling-alg.pdf}}%
    \put(0.86024605,0.07384453){\color[rgb]{0,0,0}\makebox(0,0)[lb]{\smash{\makebox[0pt]{Matrix $C$}}}}%
    \put(0,0){\includegraphics[width=\unitlength,page=4]{tiling-alg.pdf}}%
    \put(0.04068672,0.04074384){\color[rgb]{0,0,0}\makebox(0,0)[lb]{\smash{Outer loop over tiles}}}%
    \put(0,0){\includegraphics[width=\unitlength,page=5]{tiling-alg.pdf}}%
    \put(0.51048331,0.4118783){\color[rgb]{0,0,0}\makebox(0,0)[lb]{\smash{\makebox[0pt]{tile size $T$}}}}%
    \put(0,0){\includegraphics[width=\unitlength,page=6]{tiling-alg.pdf}}%
    \put(0.04074621,0.00719233){\color[rgb]{0,0,0}\makebox(0,0)[lb]{\smash{Current tile in outer loop}}}%
    \put(0,0){\includegraphics[width=\unitlength,page=7]{tiling-alg.pdf}}%
    \put(0.39497566,0.04092193){\color[rgb]{0,0,0}\makebox(0,0)[lb]{\smash{Inner loop over elements}}}%
    \put(0,0){\includegraphics[width=\unitlength,page=8]{tiling-alg.pdf}}%
    \put(0.39502797,0.00721001){\color[rgb]{0,0,0}\makebox(0,0)[lb]{\smash{Current element in inner loop}}}%
    \put(0,0){\includegraphics[width=\unitlength,page=9]{tiling-alg.pdf}}%
    \put(0.85902713,0.41521687){\color[rgb]{0,0,0}\makebox(0,0)[lb]{\smash{\makebox[0pt]{matrix size $N$}}}}%
    \put(0.75338156,0.04091382){\color[rgb]{0,0,0}\makebox(0,0)[lb]{\smash{Temporary result tile}}}%
    \put(0,0){\includegraphics[width=\unitlength,page=10]{tiling-alg.pdf}}%
  \end{picture}%
\endgroup%

%% file: content/results.tex
\section{Parameter Tuning}

\input{content/gpus_autotuning.tex}

As hyperthreading is deactivated for the Haswell CPU and as we have found an efficient number of threads $e=16^2$ for Nvidia GPUs in previous work, only the tile size $T$ is used for tuning for these architectures, see Fig.~\ref{fig:results_gpus}. An obvious observation for Haswell is that doubling the tile size often also doubles the achieved performance, while $T=4$ seems to be optimal for current GPU generations.

\input{content/knl_autotuning.tex}

Tuning for KNL and Power8 adds the number of hardware threads as a second parameter, see Fig.~\ref{fig:results_knl} for KNL. We see that optimal parameter combinations highly depend on the chosen precision and compiler. The double precision binary created by the Intel compiler using a single hardware thread results in best performance of 510 \GFLOPS{}.
We also do a measurement for the KNL in flat memory mode directly using the MCDRAM instead of the caching mechanism. Except for a slightly better performance ($\sim 2 \%$), the results are the same.

For Power8 we test from $T=16$ up to $T=512$ and from one to eight hardware threads always using only powers of two as parameters similar to KNL (not shown). Contrary to KNL, optimization for the Power8 architecture deliver similar performance results for a variety of parameters even when using the IBM XL compiler.
We don't see large deviations from our tuning results for the control case $N=7168$ (not shown) on all architectures. Although bigger matrix sizes improve the \GFLOPS{} slightly, optimum parameters remain the same.

\begin{table}[t]
\centering
\bgroup
\def\arraystretch{1.1}%
\setlength\tabcolsep{2pt}
\scriptsize
\begin{tabular}{ccccccccc}
\hline
\multirow{2}{*}{Architecture} & \multirow{2}{*}{Compiler} & Preci- & HW & Optimized & $\text{K}(S,T)$ & \multicolumn{3}{c}{Cache per HW thread} \\ \hhline{~~~~~~---}
& & sion & Threads & tile size $T$ & (see (\ref{eq:tilesize})) & L1 & L2 & L3 \\
\hline \hline
\multirow{2}{*}{P100 (nvlink)} & \multirow{6}{*}{CUDA} & single & \multirow{6}{*}{--} & 4 & 128 B & \multirow{6}{*}{--} & \multirow{6}{*}{--} & \multirow{6}{*}{--} \\ \hhline{~~-~--~~~}
 &  & double & & 4 & 256 B & & & \\ \hhline{-~-~--~~~}
\multirow{2}{*}{P100 (pci)} &  & single &  & 4 & 128 B & & & \\ \hhline{~~-~--~~~}
 &  & double & & 4 & 256 B & & & \\ \hhline{-~-~--~~~}
\multirow{2}{*}{K80} &  & single &  & 4 & 128 B & & &  \\ \hhline{~~-~--~~~}
 &  & double & & 2 & 64 B & & & \\
\hline
\multirow{4}{*}{Haswell} & \multirow{2}{*}{Intel} & single & \multirow{4}{*}{1} & 64 & \cellcolor{clr_tbl_bg1}32 KB & \cellcolor{clr_tbl_bg1}64 KB & 256 KB & $2.5$ MB \\ \hhline{~~-~-----}
 &  & double &  & 128 & \cellcolor{clr_tbl_bg2}256 KB & 64 KB & \cellcolor{clr_tbl_bg2}256 KB & $2.5$ MB \\ \hhline{~--~-----}
 & \multirow{2}{*}{GNU} & single & & 128 &\cellcolor{clr_tbl_bg1}128 KB & 64 KB & \cellcolor{clr_tbl_bg1}256 KB & $2.5$ MB \\ \hhline{~~-~-----}
 &  & double & & 128 & \cellcolor{clr_tbl_bg2}256 KB & 64 KB & \cellcolor{clr_tbl_bg2}256 KB & $2.5$ MB \\ 
\hline
\multirow{4}{*}{KNL} & \multirow{2}{*}{Intel} & single & 2 & 64 & \cellcolor{clr_tbl_bg1}32 KB & \cellcolor{clr_tbl_bg1}32 KB &  256 KB & \multirow{4}{*}{--} \\ \hhline{~~~-----~}
 &  & double & 1 & 64 & \cellcolor{clr_tbl_bg2}64 KB & \cellcolor{clr_tbl_bg2}64 KB & 512 KB & \\ \hhline{~~------~}
 & \multirow{2}{*}{GNU} & single & 1 & 256 & \cellcolor{clr_tbl_bg1}512 KB & 64 KB & \cellcolor{clr_tbl_bg1}512 KB & \\ \hhline{~~~-----~}
 &  & double & 2 & 128 & \cellcolor{clr_tbl_bg2}256 KB & 32 KB & \cellcolor{clr_tbl_bg2}256 KB & \\ 
\hline 
\multirow{4}{*}{Power8} & \multirow{2}{*}{XL} & single & 2 & 512 & \cellcolor{clr_tbl_bg1}2 MB & 32 KB & 256 KB & \cellcolor{clr_tbl_bg1}4 MB \\ \hhline{~~~------}
 &  & double & 2 & 512 & \cellcolor{clr_tbl_bg2}4 MB & 32 KB & 256 KB & \cellcolor{clr_tbl_bg2}4 MB \\ \hhline{~~-------}
 & \multirow{2}{*}{GNU} & single & 8 & 256 & \cellcolor{clr_tbl_bg1}512 KB & 8 KB & 64 KB & \cellcolor{clr_tbl_bg1}1 MB \\ \hhline{~~~------}
 &  & double & 4 & 256 & \cellcolor{clr_tbl_bg2}1 MB & 16 KB & 128 KB & \cellcolor{clr_tbl_bg2}1 MB \\ 
\hline 
\end{tabular}
\egroup
\vspace{2pt}
\caption{Estimated optimal tile size $T$ and number of hardware (HW) threads. Memory for $A$ and $B$ tiles $\text{K}(S,T)$ (Eq. \ref{eq:tilesize}) and the available cache per HW thread and cache level are listed in addition. The first cache level that can hold a complete tile is marked.
}
\label{tab:autotuning}
\end{table}

Tuning results are found in Tab.~\ref{tab:autotuning}, while the corresponding mapping of Alpaka parallel hierarchies to hardware in the case of double precision and vendor compilers selected is presented in Fig.~\ref{fig:mapping}.

\input{content/mapping.tex}

\section{Results of the scaling}

\input{content/scaling.tex}

Fig. \ref{fig:scaling_double} and \ref{fig:scaling_float} show the achieved GEMM \GFLOPS{} for all architectures considered, for both double and single precision and optimum parameter sets \cite{matthes_alexander_2017_439528}. The Nvidia P100 as expected shows the best absolute performance in all cases, while the Power8 runtime is surprisingly faster than the K80 although the Nvidia GPU has a higher theoretical peak performance than the IBM CPU.
The KNL shows a drastic drop in peak performance every second or fourth measurement beginning with $N=8192$ for both precisions, regardless of using cached or flat memory when using the Intel compiler.
To investigate this issue a test with $N=8192$ is run in double precision but 91 hardware threads. With this we get 490 \GFLOPS{} instead of 303 \GFLOPS{} (64 threads), which is only 7\% less than for $N=7168$ and $N=9216$ (both 527 \GFLOPS{}).

Most architectures show an increase in the performance for higher $N$, with the exception of Intel Haswell which for single precision shows best peak performance (665 \GFLOPS{}) for $N=2048$ and afterwards decreases reaching a plateau at $~400$ \GFLOPS{}.
In contrast to our expectations, all GPUs show a better performance when using unified memory instead of device memory, especially for small $N$.

\input{content/scaling_rel.tex}

In order to compare results Fig.~\ref{fig:scaling_rel} shows the relative peak performance for the best parameter combinations for every architecture and single and double precision.
For architectures investigated in 2016~\cite{zenker2016alpaka}, we find similar or only slightly better performance. But whereas the last paper has stated a general performance around $~20 \%$ the most recent systems are now capable to reach almost $50 \%$ of the peak performance using Alpaka.

%% file: content/gpus_autotuning.tex
\begin{figure}[t]
	\begin{tikzpicture}
	\begin{loglogaxis}[
		ybar,
		title={Tuning for P100 and K80},
		title style={font=\scriptsize},
		width=0.56\textwidth,
		height=0.335\textheight,
		xlabel={tile size $T$},
		x label style={at={(0.5,0.03)}},
		ylabel={achieved \GFLOPS},
		ylabel style={align=center,font=\scriptsize},
		yticklabel style={font=\scriptsize},
		y label style={at={(0.08,0.5)}},
		xlabel style={align=center,font=\scriptsize},
		xticklabel style={font=\scriptsize},
		legend entries={{P100 (float,nvlink)},{P100 (double,nvlink)},{P100 (float,pcie)},{P100 (double,pcie)},{K80 (float)},{K80 (double)}},
		xtick={1,2,4,8},
		xticklabels={1,2,4,8},
		ymajorgrids={true},
		major grid style={dotted},
		ytick={50,100,200,500,1000,2000,5000},
		yticklabels={50,100,200,500,1000,2000,5000},
		legend style={at={(0.885,0.52)}},
		every node near coord/.append style={anchor=west,font=\scriptsize},
		point meta=explicit symbolic,
		ymax=6500,
		ymin=4,
		xmin=0.67,
		xmax=11.9,
		legend style={fill opacity=0.92,text opacity=1,draw opacity=1,font=\scriptsize},
		bar width=3.5pt,
	]
	
	\addplot[black,fill=clr_p100] table[x=tilesize,y=P100_10,meta=P100_10] {./results/gpus_float.csv};
	\addplot[black,fill=clr_p100,pattern color=white,postaction={pattern=north east lines}] table[x=tilesize,y=P100_10,meta=P100_10] {./results/gpus.csv};

	\addplot[black,fill=clr_p100_pcie] table[x=tilesize,y=P100_pcie_10,meta=P100_pcie_10] {./results/gpus_float.csv};
	\addplot[black,fill=clr_p100_pcie,pattern color=white,postaction={pattern=north east lines}] table[x=tilesize,y=P100_pcie_10,meta=P100_pcie_10] {./results/gpus.csv};

	\addplot[black,fill=clr_k80] table[x=tilesize,y=K80_10,meta=K80_10] {./results/gpus_float.csv};
	\addplot[black,fill=clr_k80,pattern color=white,postaction={pattern=north east lines}] table[x=tilesize,y=K80_10,meta=K80_10] {./results/gpus.csv};

	\end{loglogaxis}
	\end{tikzpicture}
	\hfill
	\begin{tikzpicture}
	\begin{loglogaxis}[
		title={Tuning for Haswell},
		ybar,
		width=0.52\textwidth,
		height=0.335\textheight,
		xlabel={tile size $T$},
		ylabel={achieved \GFLOPS},
		x label style={at={(0.5,0.03)}},
		ylabel={achieved \GFLOPS},
		ylabel style={align=center,font=\scriptsize},
		yticklabel style={font=\scriptsize},
		y label style={at={(0.1,0.5)}},
		xlabel style={align=center,font=\scriptsize},
		xticklabel style={font=\scriptsize},
		legend entries={{Haswell (float,icc)},{Haswell (float,gcc)},{Haswell (double,icc)},{Haswell (double,gcc)}},
		xtick={16,32,64,128,256,512},
		xticklabels={{16},{32},{64},{128},{256},{512}},
		ymajorgrids={true},
		major grid style={dotted},
		ytick={50,100,200,500},
		yticklabels={50,100,200,500},
		xmin = 10.5,
		xmax = 390,
		ymin = 25,
		ymax = 560,
		legend style={at={(0.935,0.36)}},
		mark options={solid},
		legend style={fill opacity=0.92,text opacity=1,draw opacity=1,font=\scriptsize},
		title style={font=\scriptsize},
		point meta=explicit symbolic,
		bar width=3.5pt,
	]

	\addplot[black,fill=clr_haswell] table[x=tilesize,y=Haswell_icc_float_10,meta=Haswell_icc_float_10] {./results/cpus.csv};
	\addplot[black,fill=clr_haswell,pattern color=white,postaction={pattern=north east lines}] table[x=tilesize,y=Haswell_gcc_float_10,meta=Haswell_gcc_float_10] {./results/cpus.csv};
	\addplot[black,fill=clr_haswell_double] table[x=tilesize,y=Haswell_icc_10,meta=Haswell_icc_10] {./results/cpus.csv};
	\addplot[black,fill=clr_haswell_double,pattern color=white,postaction={pattern=north east lines}] table[x=tilesize,y=Haswell_gcc_10,meta=Haswell_gcc_10] {./results/cpus.csv};

	\end{loglogaxis}
	\end{tikzpicture}
	\caption{Achievable \GFLOPS{} for Nvidia K80 and P100, and for Intel Haswell depending on the compiler, the floating point precision and the chosen tile size of the GEMM algorithm. As there are not lesser cores than hardware threads, all of them are used.}
	\label{fig:results_gpus}
\end{figure}
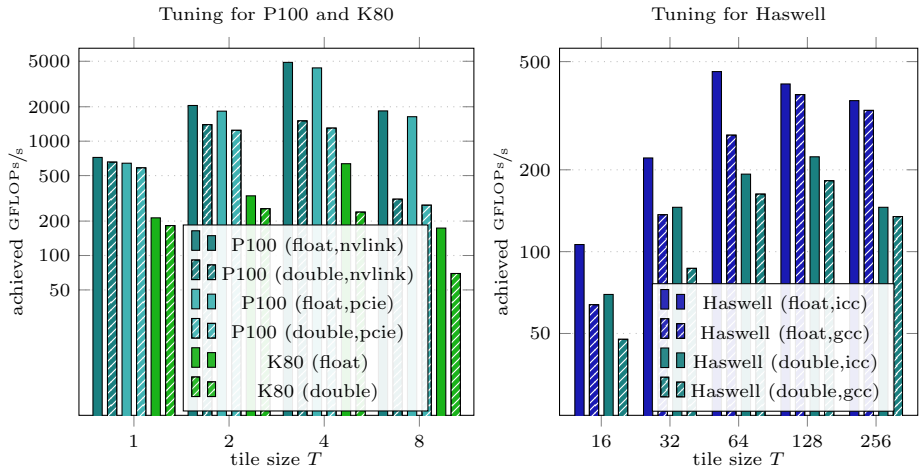

%% file: content/knl_autotuning.tex
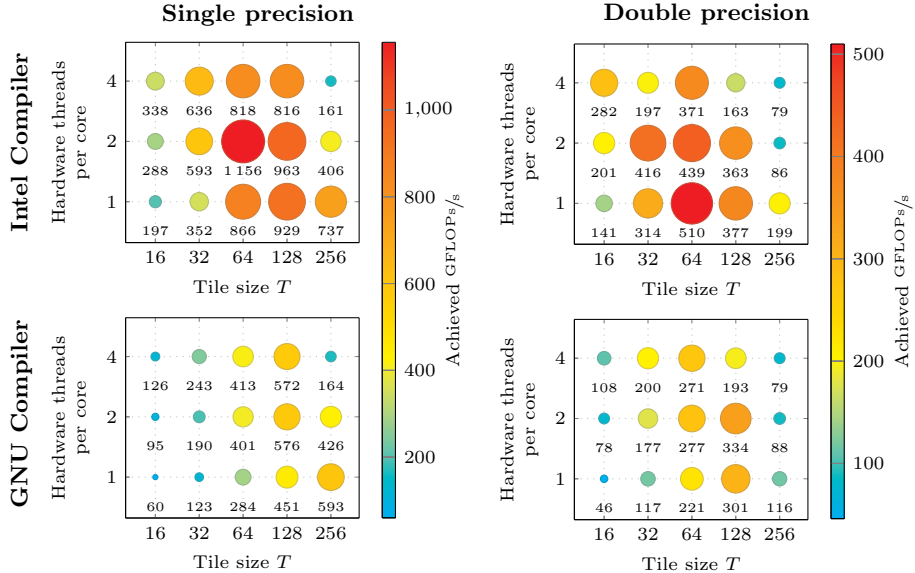
\begin{figure}[t]
\centering
\begin{minipage}{0.02\textwidth}
\rotatebox{90}{\hspace*{5pt} \textbf{GNU Compiler \hspace*{26pt} Intel Compiler}}
\end{minipage}
\begin{minipage}{0.97\textwidth}
\begin{minipage}{0.495\textwidth}
\centering
\textbf{Single precision}

\vspace*{5pt}

\begin{tikzpicture}
	\begin{loglogaxis}[
		name=plot1,
		width=0.8\textwidth,
		height=0.22\textheight,
		major tick length={1.5pt},
		xlabel={Tile size $T$},
		xtick={16,32,64,128,256},
		xticklabels={16,32,64,128,256},
		xlabel style={font=\scriptsize},
		xticklabel style={font=\scriptsize},
		ylabel={Hardware threads\\per core},
		ytick={64,128,256},
		yticklabels={1,2,4},
		ylabel style={align=center,font=\scriptsize},
		yticklabel style={font=\scriptsize},
		zlabel={\GFLOPS},
		grid=major,
		xmin = 10,
		xmax = 400,
		ymin = 40,
		ymax = 400,
		major grid style={dotted},
		colorbar,
		colorbar/width=5pt,
		colormap name=mycolormap,
		colorbar style={ylabel style={font=\scriptsize},ylabel={Achieved \GFLOPS{}},yticklabel style={font=\scriptsize,text width=1em},height=180pt},
		view={0}{90},
		point meta min=60,
		point meta max=1156
	]

	\addplot3[ %
	    scatter=true,
	    only marks,
	    mark=*,
	    color=black,
	    point meta=explicit,
	    visualization depends on = {z/19 \as \perpointmarksize},
	    scatter/@pre marker code/.append style={/tikz/mark size=\perpointmarksize},
	    visualization depends on={meta \as \nodedescription},
	    nodes near coords*={\numprint{\nodedescription}},
		every node near coord/.append style={xshift=0pt,yshift=-16pt,font=\tiny},
	]  table[x=tilesize,y=core_count,z=knl_icc_float_size_10,meta=knl_icc_float_10] {./results/knl_autotuning.csv};

	\end{loglogaxis}

	\begin{loglogaxis}[
		at={($(plot1.south)-(0,1cm)$)},
		anchor=north,
		width=0.8\textwidth,
		height=0.22\textheight,
		major tick length={1.5pt},
		xlabel={Tile size $T$},
		xtick={16,32,64,128,256},
		xticklabels={16,32,64,128,256},
		xlabel style={font=\scriptsize},
		xticklabel style={font=\scriptsize},
		ylabel={Hardware threads\\per core},
		ytick={64,128,256},
		yticklabels={1,2,4},
		ylabel style={align=center,font=\scriptsize},
		yticklabel style={font=\scriptsize},
		zlabel={\GFLOPS},
		grid=major,
		xmin = 10,
		xmax = 400,
		ymin = 40,
		ymax = 400,
		major grid style={dotted},
		view={0}{90},
		point meta min=60,
		point meta max=1156
	]

	\addplot3[ %
	    scatter=true,
	    only marks,
	    mark=*,
	    color=black,
	    point meta=explicit,
	    visualization depends on = {z/19 \as \perpointmarksize},
	    scatter/@pre marker code/.append style={/tikz/mark size=\perpointmarksize},
	    visualization depends on={meta \as \nodedescription},
	    nodes near coords*={\numprint{\nodedescription}},
		every node near coord/.append style={xshift=0pt,yshift=-16pt,font=\tiny},
	]  table[x=tilesize,y=core_count,z=knl_gcc_float_size_10,meta=knl_gcc_float_10] {./results/knl_autotuning.csv};

	\end{loglogaxis}
\end{tikzpicture}
\end{minipage}
\begin{minipage}{0.495\textwidth}
\centering
\textbf{Double precision}

\vspace*{5pt}

\begin{tikzpicture}
	\begin{loglogaxis}[
		name=plot3,
		width=0.8\textwidth,
		height=0.22\textheight,
		major tick length={1.5pt},
		xlabel={Tile size $T$},
		xtick={16,32,64,128,256},
		xticklabels={16,32,64,128,256},
		xlabel style={font=\scriptsize},
		xticklabel style={font=\scriptsize},
		ylabel={Hardware threads\\per core},
		ytick={64,128,256},
		yticklabels={1,2,4},
		ylabel style={align=center,font=\scriptsize},
		yticklabel style={font=\scriptsize},
		zlabel={\GFLOPS},
		grid=major,
		xmin = 10,
		xmax = 400,
		ymin = 40,
		ymax = 400,
		major grid style={dotted},
		colorbar,
		colorbar/width=5pt,
		colormap name=mycolormap,
		colorbar style={ylabel style={font=\scriptsize},ylabel={Achieved \GFLOPS{}},yticklabel style={font=\scriptsize,text width=1em},height=180pt},
		view={0}{90},
		point meta min=46,
		point meta max=510
	]

	\addplot3[ %
	    scatter=true,
	    only marks,
	    mark=*,
	    color=black,
	    point meta=explicit,
	    visualization depends on = {z/11 \as \perpointmarksize},
	    scatter/@pre marker code/.append style={/tikz/mark size=\perpointmarksize},
	    visualization depends on={meta \as \nodedescription},
	    nodes near coords*={\numprint{\nodedescription}},
		every node near coord/.append style={xshift=0pt,yshift=-16pt,font=\tiny},
	]  table[x=tilesize,y=core_count,z=knl_icc_size_10,meta=knl_icc_10] {./results/knl_autotuning.csv};

	\end{loglogaxis}
	
	\begin{loglogaxis}[
		at={($(plot1.south)-(0,1cm)$)},
		anchor=north,
		width=0.8\textwidth,
		height=0.22\textheight,
		major tick length={1.5pt},
		xlabel={Tile size $T$},
		xtick={16,32,64,128,256},
		xticklabels={16,32,64,128,256},
		xlabel style={font=\scriptsize},
		xticklabel style={font=\scriptsize},
		ylabel={Hardware threads\\per core},
		ytick={64,128,256},
		yticklabels={1,2,4},
		ylabel style={align=center,font=\scriptsize},
		yticklabel style={font=\scriptsize},
		zlabel={\GFLOPS},
		grid=major,
		xmin = 10,
		xmax = 400,
		ymin = 40,
		ymax = 400,
		major grid style={dotted},
		view={0}{90},
		point meta min=46,
		point meta max=510
	]

	\addplot3[ %
	    scatter=true,
	    only marks,
	    mark=*,
	    color=black,
	    point meta=explicit,
	    visualization depends on = {z/11 \as \perpointmarksize},
	    scatter/@pre marker code/.append style={/tikz/mark size=\perpointmarksize},
	    visualization depends on={meta \as \nodedescription},
	    nodes near coords*={\numprint{\nodedescription}},
		every node near coord/.append style={xshift=0pt,yshift=-16pt,font=\tiny},
	]  table[x=tilesize,y=core_count,z=knl_gcc_size_10,meta=knl_gcc_10] {./results/knl_autotuning.csv};

	\end{loglogaxis}

\end{tikzpicture}
\end{minipage}
\end{minipage}
	\caption{Achievable \GFLOPS{} for Intel Xeon Phi Knights Landing (KNL) depending on the compiler, the floating point precision, the chosen tile size of the tiled matrix multiplication algorithm and the used hardware threads per core. The bigger the mark size the higher the achieved \GFLOPS{}. The mark radius is calculated with ${(\text{achieved }\GFLOPSinner{})}^{\sfrac{5}{7}}$ as this has been shown a good value for human perception~\cite{Gum11}. GNU compiler 6.2 and Intel compiler 17 are used. For compiler options see Tab.~\ref{tab:compilers}.}
	\label{fig:results_knl}
\end{figure}

%% file: content/mapping.tex
\begin{figure}[!tbp]
  \centering
  \includegraphics[width=0.8\textwidth]{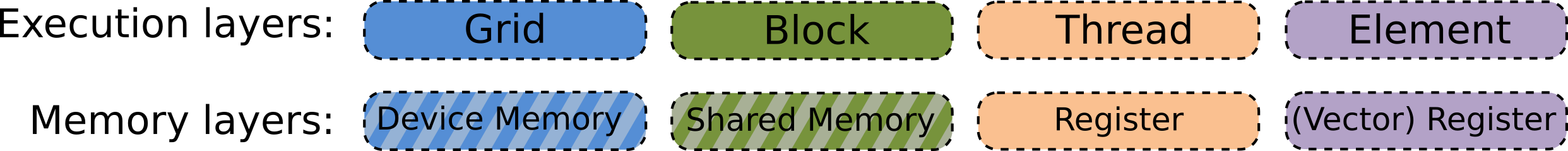}
  \hfill
  \begin{flushleft}
  \begin{minipage}[b]{0.5395\textwidth}
    \centering
    \includegraphics[width=\textwidth]{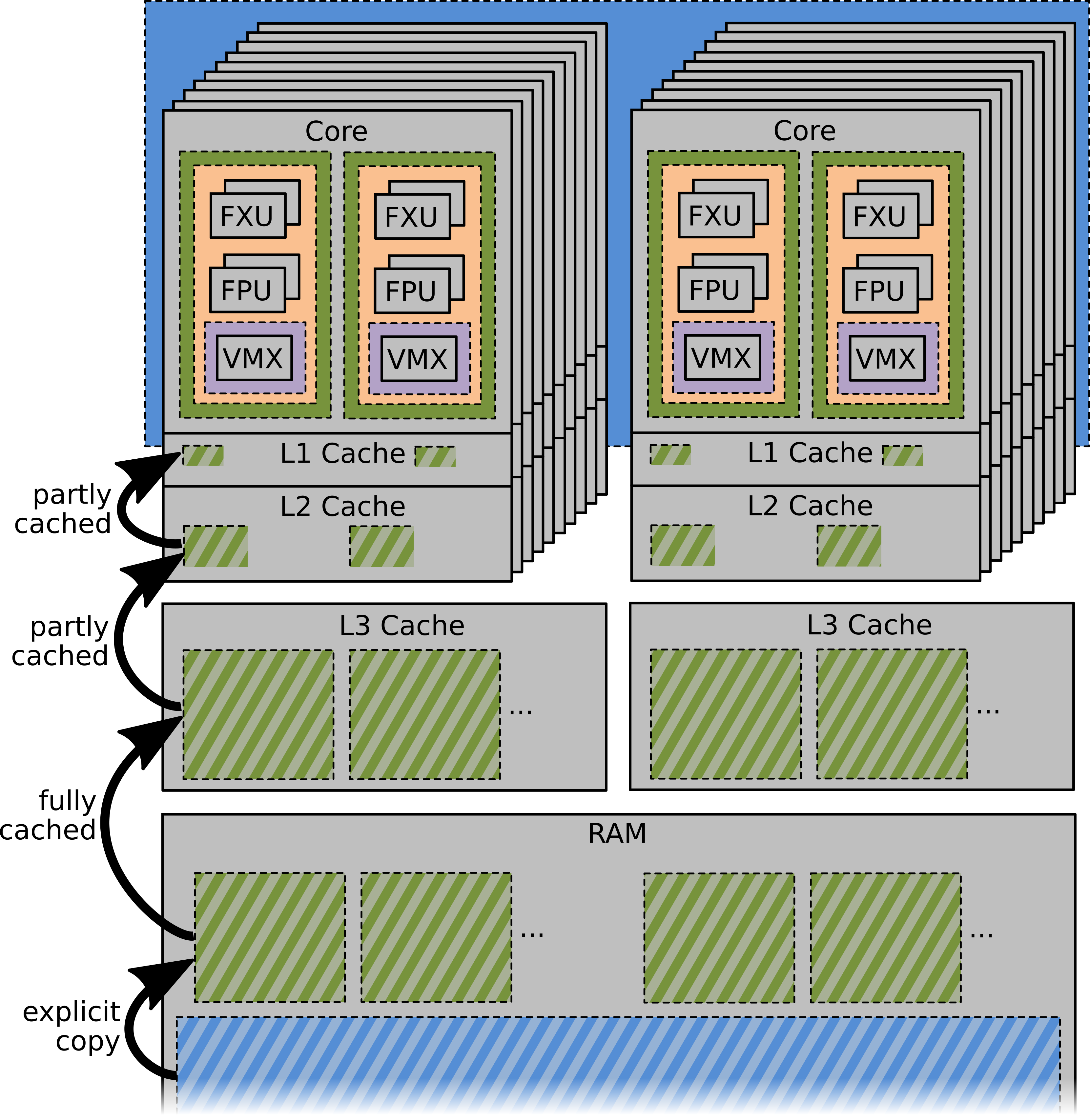}
  \end{minipage}
  \hfill
  \begin{minipage}[b]{0.4405\textwidth}
    \centering
    \includegraphics[width=\textwidth]{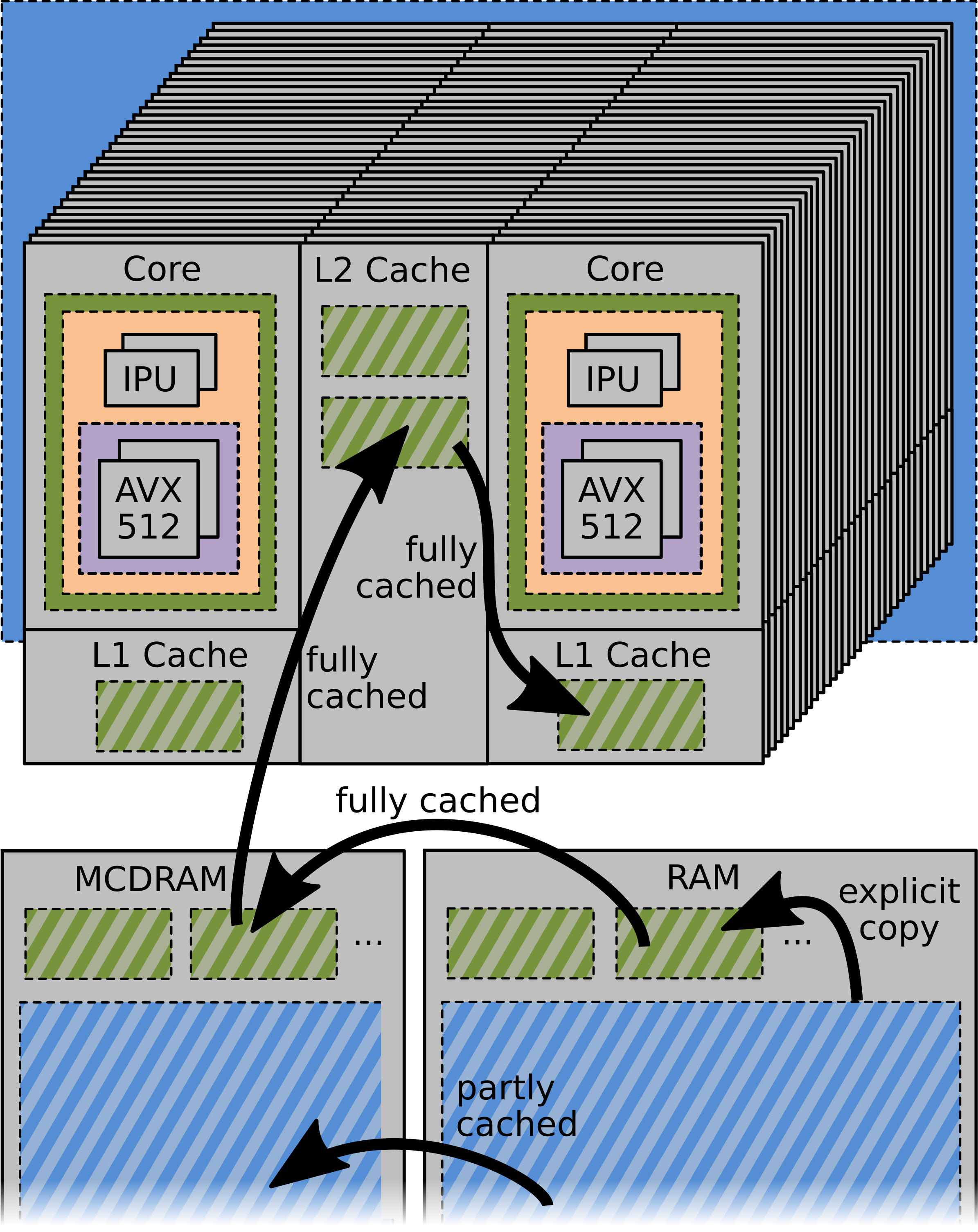}
  \end{minipage}
  \hfill
  \begin{minipage}[b]{0.54\textwidth}
	\centering
    IBM Power8
  \end{minipage}
  \hfill
  \begin{minipage}[b]{0.44\textwidth}
	\centering
    Intel KNL
  \end{minipage}
  \end{flushleft}
  \hfill
  \begin{minipage}[b]{1\textwidth}
    \centering
    \includegraphics[width=\textwidth]{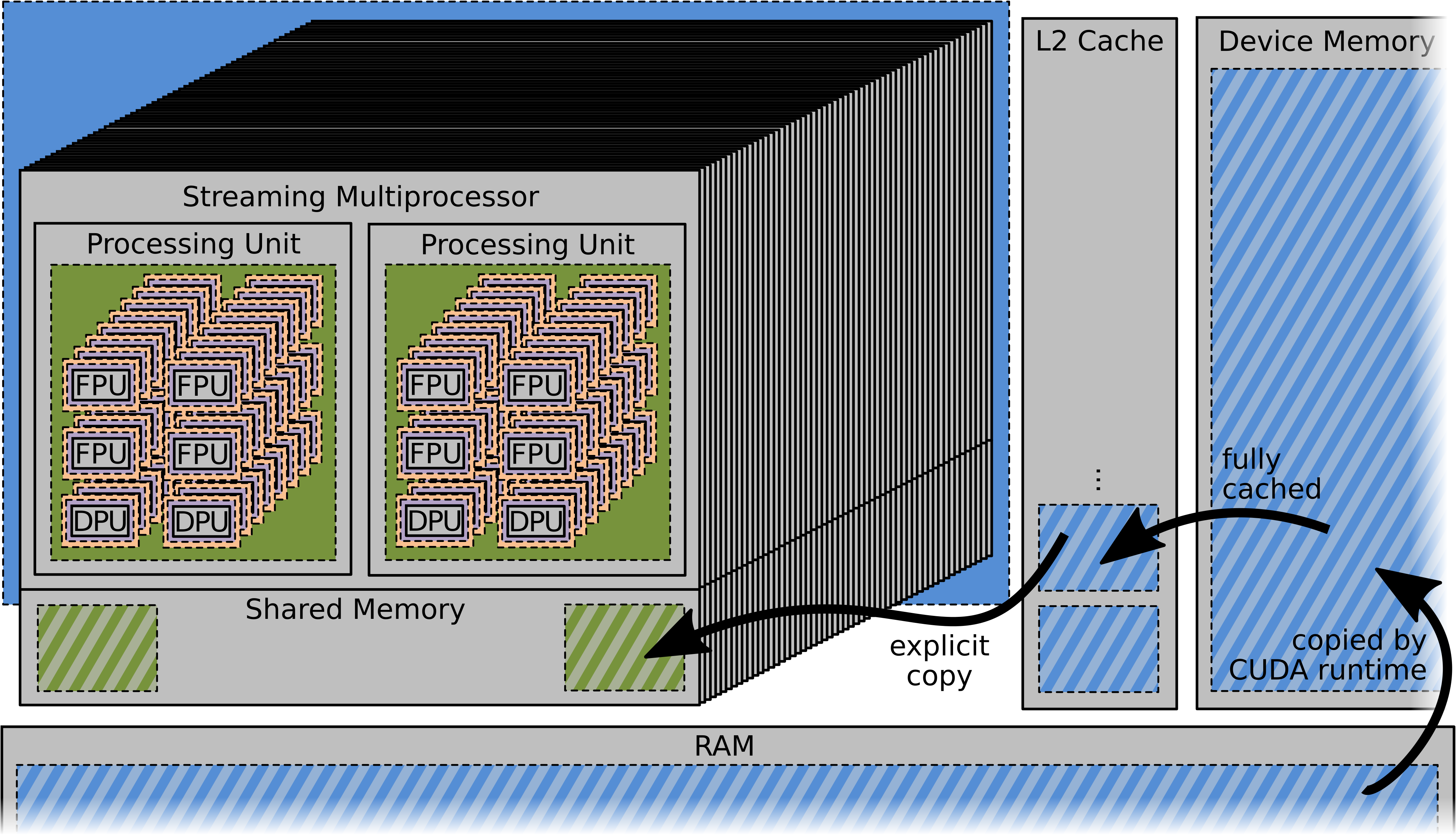}
    Nvidia Tesla P100
  \end{minipage}
  \caption {Alpaka mappings for IBM's Power8, Intel's KNL, and Nvidia's Tesla P100. Every mapping uses the optimal parameters of the parameter tuning for double precision and the vendor compiler from Tab. \ref{tab:autotuning}. The CPU mappings use the OpenMP2 Block back end. The GPU mapping uses the CUDA back end and unified memory.}
  \label{fig:mapping}
\end{figure}

%% file: content/scaling.tex
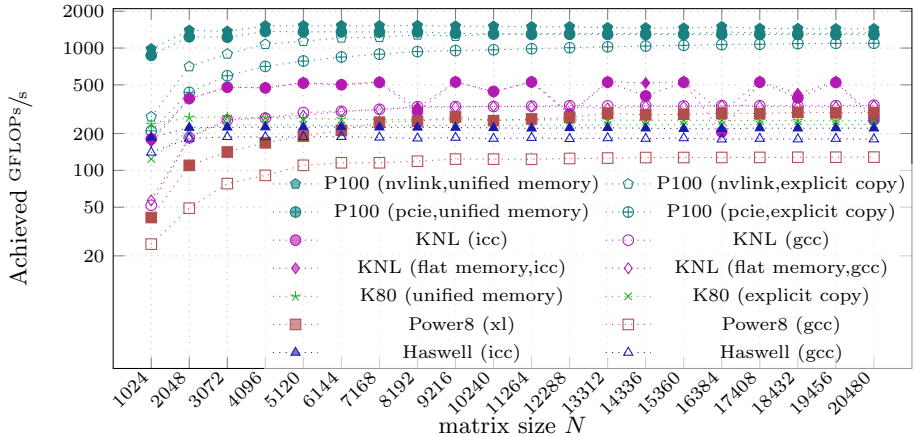
\begin{figure}[t]
	\pgfplotsset{
	    layers/legend behind plots/.define layer set={
	            axis background,axis grid,axis ticks,axis lines,axis tick labels,main,axis descriptions,axis foreground
	    }{
	        grid style= {/pgfplots/on layer=axis grid},
	        tick style= {/pgfplots/on layer=axis ticks},
	        axis line style= {/pgfplots/on layer=axis lines},
	        label style= {/pgfplots/on layer=axis descriptions},
	        legend style= {/pgfplots/on layer=axis tick labels}, 
	        title style= {/pgfplots/on layer=axis descriptions},
	        colorbar style= {/pgfplots/on layer=axis descriptions},
	        ticklabel style= {/pgfplots/on layer=axis tick labels},
	        axis background@ style={/pgfplots/on layer=axis background},
	        3d box foreground style={/pgfplots/on layer=axis foreground},
	    },
	}
	\begin{tikzpicture}
	\begin{semilogyaxis}[
		width=1\textwidth,
		height=0.33\textheight,
		xlabel={matrix size $N$},
		ylabel={Achieved \GFLOPS{}},
        legend columns=2,
		legend style={draw=none,fill opacity=0.6,text opacity=1,draw opacity=1,font=\scriptsize},
		set layers=legend behind plots, 
		cell picture=true, 
   		legend entries={
	   		{P100 (nvlink,unified memory)},{P100 (nvlink,explicit copy)},{P100 (pcie,unified memory)},{P100 (pcie,explicit copy)},
	   		{KNL (icc)},{KNL (gcc)},{KNL (flat memory,icc)},{KNL (flat memory,gcc)},
	   		{K80 (unified memory)},{K80 (explicit copy)},
	   		{Power8 (xl)},{Power8 (gcc)},
	   		{Haswell (icc)}, {Haswell (gcc)},
	   	},
		xtick={1024,2048,3072,4096,5120,6144,7168,8192,9216,10240,11264,12288,13312,14336,15360,16384,17408,18432,19456,20480},
		xticklabels={1024,2048,3072,4096,5120,6144,7168,8192,9216,10240,11264,12288,13312,14336,15360,16384,17408,18432,19456,20480},
		x tick label style={rotate=45,anchor=east},
		every tick label/.append style={font=\scriptsize},
		scaled x ticks = false,
		xmajorgrids={true},
		major grid style={dotted},
		ytick={20,50,100,200,500,1000,2000},
		yticklabels={20,50,100,200,500,1000,2000},
		grid = both,
		xmin = 0,
		xmax = 21504,
		ymin = 2.4,
		ymax = 2100,
		legend style={at={(1.01,.568)}},
		every node near coord/.append style={anchor=west},
		mark options={solid},
		point meta=explicit symbolic,
	]
	

	\addplot[mark=pentagon*,clr_p100,line width=0.4,dotted] table[x=kmn,y=P100_power8,meta=P100_power8] {./results/big_scaling.csv};
	\addplot[mark=pentagon,clr_p100,line width=0.4,dotted] table[x=kmn,y=P100_power8_memcpy,meta=P100_power8_memcpy] {./results/big_scaling.csv};
	\addplot[mark=oplus*,clr_p100,line width=0.4,dotted] table[x=kmn,y=P100_pcie,meta=P100_pcie] {./results/big_scaling.csv};
	\addplot[mark=oplus,clr_p100,line width=0.4,dotted] table[x=kmn,y=P100_pcie_memcpy,meta=P100_pcie_memcpy] {./results/big_scaling.csv};

	\addplot[mark=*,clr_knl,line width=0.4,dotted] table[x=kmn,y=KNL_icc,meta=KNL_icc] {./results/big_scaling.csv};
	\addplot[mark=o,clr_knl,line width=0.4,dotted] table[x=kmn,y=KNL_gcc,meta=KNL_gcc] {./results/big_scaling.csv};
	\addplot[mark=diamond*,clr_knl,line width=0.4,dotted] table[x=kmn,y=KNL_icc_flat,meta=KNL_icc_flat] {./results/big_scaling.csv};
	\addplot[mark=diamond,clr_knl,line width=0.4,dotted] table[x=kmn,y=KNL_gcc_flat,meta=KNL_gcc_flat] {./results/big_scaling.csv};

	\addplot[mark=star,clr_k80,line width=0.4,dotted] table[x=kmn,y=K80,meta=K80] {./results/big_scaling.csv};
	\addplot[mark=x,clr_k80,line width=0.4,dotted] table[x=kmn,y=K80_memcpy,meta=K80_memcpy] {./results/big_scaling.csv};
		
	\addplot[mark=square*,clr_power8,line width=0.4,dotted] table[x=kmn,y=Power8_xl,meta=Power8_xl] {./results/big_scaling.csv};
	\addplot[mark=square,clr_power8,line width=0.4,dotted] table[x=kmn,y=Power8_gcc,meta=Power8_gcc] {./results/big_scaling.csv};

	\addplot[mark=triangle*,clr_haswell,line width=0.4,dotted] table[x=kmn,y=Haswell_icc,meta=Haswell_icc] {./results/big_scaling.csv};
	\addplot[mark=triangle,clr_haswell,line width=0.4,dotted] table[x=kmn,y=Haswell_gcc,meta=Haswell_gcc] {./results/big_scaling.csv};

	
	\end{semilogyaxis}
	\end{tikzpicture}
	\caption{Achievable \GFLOPS{} for all considered architectures for double precision depending on the matrix size and the compiler.}
	\label{fig:scaling_double}
\end{figure}

\begin{figure}[t]
	\begin{tikzpicture}
	\pgfplotsset{
	    layers/legend behind plots/.define layer set={
	            axis background,axis grid,axis ticks,axis lines,axis tick labels,main,axis descriptions,axis foreground
	    }{
	        grid style= {/pgfplots/on layer=axis grid},
	        tick style= {/pgfplots/on layer=axis ticks},
	        axis line style= {/pgfplots/on layer=axis lines},
	        label style= {/pgfplots/on layer=axis descriptions},
	        legend style= {/pgfplots/on layer=axis tick labels}, 
	        title style= {/pgfplots/on layer=axis descriptions},
	        colorbar style= {/pgfplots/on layer=axis descriptions},
	        ticklabel style= {/pgfplots/on layer=axis tick labels},
	        axis background@ style={/pgfplots/on layer=axis background},
	        3d box foreground style={/pgfplots/on layer=axis foreground},
	    },
	}
	\begin{semilogyaxis}[
		width=1\textwidth,
		height=0.33\textheight,
		xlabel={matrix size $N$},
		ylabel={Achieved \GFLOPS{}},
        legend columns=2,
		legend style={draw=none,fill opacity=0.6,text opacity=1,draw opacity=1,font=\scriptsize},
		set layers=legend behind plots, 
		cell picture=true, 
   		legend entries={
	   		{P100 (nvlink,unified memory)},{P100 (nvlink,explicit copy)},{P100 (pcie,unified memory)},{P100 (pcie,explicit copy)},
	   		{KNL (icc)},{KNL (gcc)},{KNL (flat memory,icc)},{KNL (flat memory,gcc)},
	   		{K80 (unified memory)},{K80 (explicit copy)},
	   		{Power8 (xl)},{Power8 (gcc)},
	   		{Haswell (icc)}, {Haswell (gcc)},
	   	},
		xtick={1024,2048,3072,4096,5120,6144,7168,8192,9216,10240,11264,12288,13312,14336,15360,16384,17408,18432,19456,20480},
		xticklabels={1024,2048,3072,4096,5120,6144,7168,8192,9216,10240,11264,12288,13312,14336,15360,16384,17408,18432,19456,20480},
		x tick label style={rotate=45,anchor=east},
		every tick label/.append style={font=\scriptsize},
		scaled x ticks = false,
		xmajorgrids={true},
		major grid style={dotted},
		ytick={20,50,100,200,500,1000,2000},
		yticklabels={20,50,100,200,500,1000,2000},
		grid = both,
		xmin = 0,
		xmax = 21504,
		ymin = 2,
		ymax = 6500,
		legend style={at={(1.01,.568)}},
		every node near coord/.append style={anchor=west},
		mark options={solid},
		point meta=explicit symbolic,
	]
	

	\addplot[mark=pentagon*,clr_p100,line width=0.4,dotted] table[x=kmn,y=P100_power8,meta=P100_power8] {./results/big_scaling_float.csv};
	\addplot[mark=pentagon,clr_p100,line width=0.4,dotted] table[x=kmn,y=P100_power8_memcpy,meta=P100_power8_memcpy] {./results/big_scaling_float.csv};
	\addplot[mark=oplus*,clr_p100,line width=0.4,dotted] table[x=kmn,y=P100_pcie,meta=P100_pcie] {./results/big_scaling_float.csv};
	\addplot[mark=oplus,clr_p100,line width=0.4,dotted] table[x=kmn,y=P100_pcie_memcpy,meta=P100_pcie_memcpy] {./results/big_scaling_float.csv};

	\addplot[mark=*,clr_knl,line width=0.4,dotted] table[x=kmn,y=KNL_icc,meta=KNL_icc] {./results/big_scaling_float.csv};
	\addplot[mark=o,clr_knl,line width=0.4,dotted] table[x=kmn,y=KNL_gcc,meta=KNL_gcc] {./results/big_scaling_float.csv};
	\addplot[mark=diamond*,clr_knl,line width=0.4,dotted] table[x=kmn,y=KNL_icc_flat,meta=KNL_icc_flat] {./results/big_scaling_float.csv};
	\addplot[mark=diamond,clr_knl,line width=0.4,dotted] table[x=kmn,y=KNL_gcc_flat,meta=KNL_gcc_flat] {./results/big_scaling_float.csv};

	\addplot[mark=star,clr_k80,line width=0.4,dotted] table[x=kmn,y=K80,meta=K80] {./results/big_scaling_float.csv};
	\addplot[mark=x,clr_k80,line width=0.4,dotted] table[x=kmn,y=K80_memcpy,meta=K80_memcpy] {./results/big_scaling_float.csv};
		
	\addplot[mark=square*,clr_power8,line width=0.4,dotted] table[x=kmn,y=Power8_xl,meta=Power8_xl] {./results/big_scaling_float.csv};
	\addplot[mark=square,clr_power8,line width=0.4,dotted] table[x=kmn,y=Power8_gcc,meta=Power8_gcc] {./results/big_scaling_float.csv};

	\addplot[mark=triangle*,clr_haswell,line width=0.4,dotted] table[x=kmn,y=Haswell_icc,meta=Haswell_icc] {./results/big_scaling_float.csv};
	\addplot[mark=triangle,clr_haswell,line width=0.4,dotted] table[x=kmn,y=Haswell_gcc,meta=Haswell_gcc] {./results/big_scaling_float.csv};

	
	\end{semilogyaxis}
	\end{tikzpicture}
	\caption{Achievable \GFLOPS{} for all considered architectures for single precision depending on the matrix size and the compiler.}
	\label{fig:scaling_float}
\end{figure}
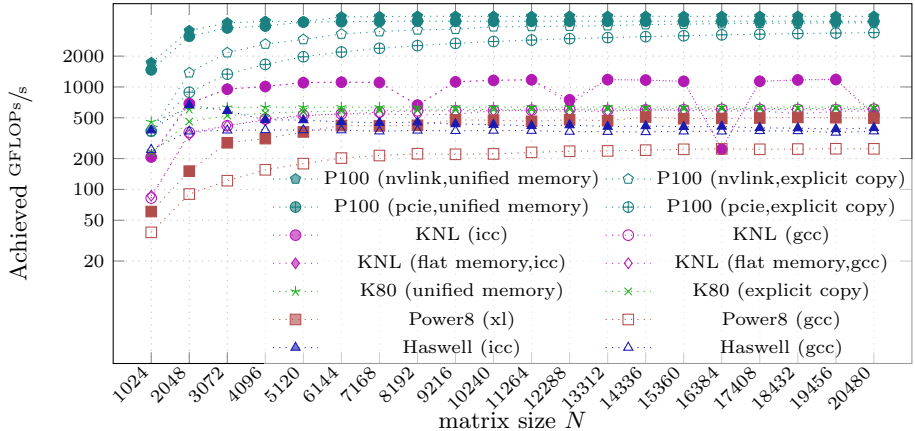

%% file: content/scaling_rel.tex
\begin{figure}[t]
	\begin{tikzpicture}
	\begin{semilogyaxis}[
		width=1\textwidth,
		height=0.4\textheight,
		xlabel={matrix size $N$},
		x label style={at={(axis description cs:0.5,-0.005)},anchor=north},
		y label style={at={(axis description cs:0.03,0.8)},anchor=east},
		ylabel={Achieved performance in \%\\relative to the peak performance},
		ylabel style={align=center},
		legend columns=2,
		legend style={fill opacity=0.6,text opacity=1,draw opacity=1,font=\scriptsize},
   		legend entries={
	   		{P100 (double,nvlink,u. memory)},{P100 (float,nvlink,u. memory)},
	   		{P100 (double,pcie,u. memory)},{P100 (float,pcie,u. memory)},
	   		{KNL (double,flat memory,icc)},{KNL (float,flat memory,icc)},
	   		{KNL (double,flat memory,gcc)},{KNL (float,flat memory,gcc)},
	   		{K80 (double,u. memory)},{K80 (float,u. memory)},
	   		{Power8 (double,xl)},{Power8 (float,xl)},
	   		{Haswell (double,icc)}, {Haswell (float,icc)},
	   	},
		xtick={1024,2048,3072,4096,5120,6144,7168,8192,9216,10240,11264,12288,13312,14336,15360,16384,17408,18432,19456,20480},
		xticklabels={1024,2048,3072,4096,5120,6144,7168,8192,9216,10240,11264,12288,13312,14336,15360,16384,17408,18432,19456,20480},
		x tick label style={rotate=45,anchor=east},
		every tick label/.append style={font=\scriptsize},
		scaled x ticks = false,
		xmajorgrids={true},
		major grid style={dotted},
		ytick={1,1.5,2,3,4,5,7,10,15,20,30,40,50},
		yticklabels={1,1.5,2,3,4,5,7,10,15,20,30,40,50},
		grid = both,
		xmin = 512,
		xmax = 20992,
		ymin = 1.4,
		ymax = 51,
		legend style={at={(0.98,.465)}},
		every node near coord/.append style={anchor=west},
		mark options={solid},
		point meta=explicit symbolic,
	]
	

	\addplot[mark=pentagon*,clr_p100,line width=0.4,dotted] table[x=kmn,y=P100_power8_rel,meta=P100_power8_rel] {./results/big_scaling.csv};
	\addplot[mark=pentagon,clr_p100,line width=1.4,dotted] table[x=kmn,y=P100_power8_rel,meta=P100_power8_rel] {./results/big_scaling_float.csv};
	\addplot[mark=oplus*,clr_p100,line width=0.4,dotted] table[x=kmn,y=P100_pcie_rel,meta=P100_pcie_rel] {./results/big_scaling.csv};
	\addplot[mark=oplus,clr_p100,line width=0.4,dotted] table[x=kmn,y=P100_pcie_rel,meta=P100_pcie_rel] {./results/big_scaling_float.csv};

	\addplot[mark=*,clr_knl,line width=0.4,dotted] table[x=kmn,y=KNL_icc_flat_rel,meta=KNL_icc_flat_rel] {./results/big_scaling.csv};
	\addplot[mark=o,clr_knl,line width=1.4,dotted] table[x=kmn,y=KNL_icc_flat_rel,meta=KNL_icc_flat_rel] {./results/big_scaling_float.csv};
	\addplot[mark=diamond*,clr_knl,line width=0.4,dotted] table[x=kmn,y=KNL_gcc_flat_rel,meta=KNL_gcc_flat_rel] {./results/big_scaling.csv};
	\addplot[mark=diamond,clr_knl,line width=0.4,dotted] table[x=kmn,y=KNL_gcc_flat_rel,meta=KNL_gcc_flat_rel] {./results/big_scaling_float.csv};

	\addplot[mark=star,clr_k80,line width=0.4,dotted] table[x=kmn,y=K80_rel,meta=K80_rel] {./results/big_scaling.csv};
	\addplot[mark=x,clr_k80,line width=0.4,dotted] table[x=kmn,y=K80_rel,meta=K80_rel] {./results/big_scaling_float.csv};
		
	\addplot[mark=square*,clr_power8,line width=1.4,dotted] table[x=kmn,y=Power8_xl_rel,meta=Power8_xl_rel] {./results/big_scaling.csv};
	\addplot[mark=square,clr_power8,line width=0.4,dotted] table[x=kmn,y=Power8_xl_rel,meta=Power8_xl_rel] {./results/big_scaling_float.csv};

	\addplot[mark=triangle*,clr_haswell,line width=0.4,dotted] table[x=kmn,y=Haswell_icc_rel,meta=Haswell_icc_rel] {./results/big_scaling.csv};
	\addplot[mark=triangle,clr_haswell,line width=1.4,dotted] table[x=kmn,y=Haswell_icc_rel,meta=Haswell_icc_rel] {./results/big_scaling_float.csv};

	
	\end{semilogyaxis}
	\end{tikzpicture}
	\caption{Achieved performances relative to the peak performance for the fastest parameter combinations of every architecture for single and double precision. Some scalings of particular interest are highlighted.}
	\label{fig:scaling_rel}
\end{figure}
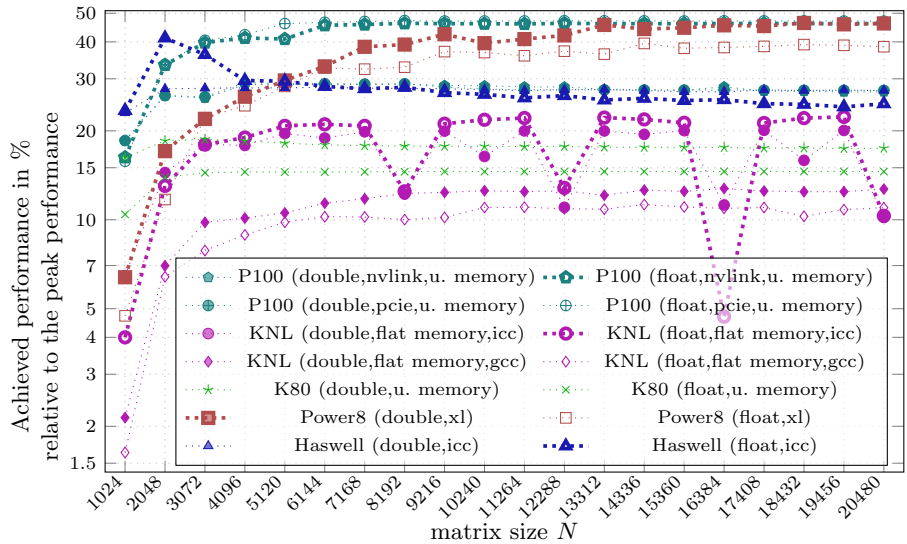

%% file: content/analysis.tex
\section{Analysis} 

\subsubsection{Autovectorization}

\begin{figure}
\begin{lstlisting}[basicstyle=\scriptsize\ttfamily,caption={Dissambled output of \texttt{objdump -DSC} for the most inner and performance critical loop of the tile matrix multiplication kernel. It shows that loop unrolling, vectorization and fused multiply add are realized by the compiler.}, label={lst:dissamble}, escapeinside={|}{|}]
for( TSize j(0); j < numElements; ++j )
{
        lineC[j] += a * lineB[j];
  422a5e:	62 a2 ed 40 b8 0c 18 	vfmadd231pd (%rax,%r11,1),%zmm18,%zmm17
  422a65:	62 a2 ed 40 b8 44 18 	vfmadd231pd 0x40(%rax,%r11,1),%zmm18,%zmm16
  422a6c:	01 
  422a6d:	62 32 ed 40 b8 7c 18 	vfmadd231pd 0x80(%rax,%r11,1),%zmm18,%zmm15
  422a74:	02 
  422a75:	62 32 ed 40 b8 74 18 	vfmadd231pd 0xc0(%rax,%r11,1),%zmm18,%zmm14
  422a7c:	03 
  422a7d:	62 32 ed 40 b8 6c 18 	vfmadd231pd 0x100(%rax,%r11,1),%zmm18,%zmm13
  422a84:	04 
  422a85:	62 32 ed 40 b8 64 18 	vfmadd231pd 0x140(%rax,%r11,1),%zmm18,%zmm12
  422a8c:	05 
  422a8d:	62 b2 ed 40 b8 64 18 	vfmadd231pd 0x180(%rax,%r11,1),%zmm18,%zmm4
  422a94:	06 
  422a95:	62 b2 ed 40 b8 5c 18 	vfmadd231pd 0x1c0(%rax,%r11,1),%zmm18,%zmm3
  422a9c:	07 
\end{lstlisting}
\tikz[overlay]{\draw[draw=blue,thick] ($(-0.05,6.485)$) rectangle ($(6.2,7.35)$);}
\tikz[overlay]{\draw ($(7.1,6.915)$) node[text=blue] {C++ code};}
\tikz[overlay]{\draw[draw=red,thick] ($(4.75,6.46)$) rectangle ($(11.75,2.5)$);}
\tikz[overlay]{\draw ($(8.25,2.3)$) node[text=red] {Unrolled assembler code};}
\tikz[overlay]{\draw[draw=olive,thick] ($(9.25,5.65)$) rectangle ($(10.25,5.35)$);}
\tikz[overlay]{\draw ($(9.75,5.15)$) node[text=olive] {AVX-512 register};}
\tikz[overlay]{\draw[draw=violet,thick] ($(4.37,4.51)$) rectangle ($(6.1,4.21)$);}
\tikz[overlay]{\draw ($(5.6,4.01)$) node[text=violet] {Fused multiply add};}
\end{figure}

Listing \ref{lst:dissamble} shows the dissembled KNL binary built by the Intel compiler for the most inner and performance critical loop of the tiling matrix multiplication kernel. C++ code is marked blue, assembler code red. With \lstinline|vfmadd231pd| being the fused multiply add vector function working on AVX-512 vectors (\lstinline|zmm*|) and loop unrolling we find that the Intel compiler is capable of optimizing the inner loop despite the heavy templated Alpaka code.

\subsubsection{Parameter tuning}

We assume that tuning for KNL resulted in best FP performance using one hardware thread (see Fig.~\ref{fig:results_knl}) because larger tiles then fit best into the L2 cache of 512 KB, which otherwise would have to be shared between threads.
This is supported by the fact that using double precision often requires smaller tile sizes than single precision. Fig.~\ref{fig:results_gpus} shows the element layer with $T=4$ causing performance gain, especially for the P100, as it has more shared memory and registers available per thread than the K80.

\subsubsection{Scaling}

Most architectures show poor performance for small matrix sizes $N\leq 2048$ which at first glance could be blamed on under-utilization, although at closer look is questionable e.g. in case of the KNL which performs $2\times 10^9$ floating point operations that clearly dominate over memory operations following Eq.~\ref{eq:R_N_T}.

We found the KNL in flat memory mode to be only about $\sim 2\%$ faster than in cached memory mode, except for very small $N$, which can be explained by the fact that the same memory is needed very often, but needs to be copied from RAM to MCDRAM only once. In all cases, using RAM only is much slower that using MCDRAM.
We see performance degradation on KNL for (almost) every second $N$ (DP) and for every fourth $N$ (SP) starting with $N=8192$, except for $N=14336$ (flat memory, DP). When choosing an uneven number of 91 hardware threads, performance improves for $N=8192$ (DP). As the issue always appears on very even numbers we assume that the KNL has performance issues if many hardware threads access the very same memory location at the same time.
As this issue does not occur for the GNU compiler, we suspect Intel's optimized OpenMP implementation to cause this.

The K80's relative peak performance is only around 15\% for single precision (SP) and around 18\% for double precision (DP) whereas the P100 reaches 46\% (SP) and 28\% (DP). As loading to shared memory is not optimally realized, we attribute this difference to the P100 having more registers per thread and more shared memory than the K80, thus more blocks can run concurrently which better hides memory latencies. Although SP values need half the space of DP the K80 has three times more SP units than DP, thus the SP version needs to load more memory for all scheduled blocks, which leads to performance degradation, which is not the case for the P100 with only two times more SP than DP units.
Another problem of the algorithmic implementation (but not of Alpaka) is that the index arithmetics lead to a unfavorable ratio of integer to floating point operations, thus degrading FPU utilization.
We emphasize that platform-dependent memory access optimizations are within the responsibility of the user code when using Alpaka.

The Haswell architecture shows a different behavior than all other systems for SP where the peak performance has its peak at $N=2048$ and then slowly decreases. For $N=2048$ matrices $A$ and $B$ use only 32~MB which fits into the L3 cache of one Haswell CPU (see Tab. 3 \ref{tab:cpus}), thus accelerating memory access.

%% file: content/evaluation.tex
\section{Conclusion}

Within the scope of this work we have shown that portable single-source C++11 code using Alpaka can run on current many-core architectures without changing any line inside the algorithmic relevant part of the source code, seeing good floating point performance for the most recent systems when reasonably designing the code for exploiting many-core parallelism.
We find that optimizing the number of hardware threads and the tile size for a simple GEMM algorithm leads to considerable increase in performance that can be well explained by the architectural characteristics and is independent of the Alpaka abstractions.

This becomes evident when analyzing the effects of vendor-specific compiler optimization. These do not only show that expected optimizations such as autovectorization, loop unrolling and the use of fused multiply adds are performed using Alpaka but that for bleeding edge hardware like Intel KNL, Nvidia P100 and IBM Power8 using vendor compilers gives a significant boost in performance.

When using vendor-specific compilers with appropriate optimization flags and \lstinline|#pragma| statements we are able to come close to 50\% of the expected peak floating point performance on the Nvidia P100 and IBM Power8 architectures, and in addition could increase the performance on well known architectures like Haswell by about five percentage points. We can thus conclude that the abstract parallel redundant hierarchy interface introduced by Alpaka does not prevent compiler optimization and tuning.
However, we also find that the performance gains observed heavily depend on the target architecture and software environment available. We express our hope that the implementation of modern C++ support in compilers relevant for high performance computing will foster the approach we take to performance portability with Alpaka.

Our analysis shows that for some architectures such as Intel's KNL more tuning parameters have to be included in order to achieve optimum results for certain problem sizes when optimizing with vendor-specific compilers. For future applications this potentially increases the time it takes for tuning a code, making tuning itself a compute- and memory-intensive task.

We clearly find that most modern vendor-specific compilers, with the prominent exception of IBM's XL compiler, are able to create highly optimized code for their target architecture from the Alpaka GEMM implementation. This shows that with Alpaka writing abstract, single-source C++ code with close-to-zero overhead is possible on todays high performance many-core architectures, demonstrating that code abstraction for sake of portability and architecture-specific tuning do not contradict each other.